\documentclass[preprint2]{aastex}

\newcommand{\nh}{N_{\rm H}}

\newcommand{\ee}{$e^\pm$}

\newcommand{\lh}{\ell_{\rm h}}
\newcommand{\ls}{\ell_{\rm s}}
\newcommand{\lth}{\ell_{\rm th}}
\newcommand{\lnth}{\ell_{\rm nth}}
\newcommand{\sax}{{\it Beppo\-SAX}}
\newcommand{\gmin}{\gamma_{\rm min}}

\newcommand{\xte}{{\it RXTE}}

\newcommand{\gro}{{\it CGRO}}
\newcommand{\cnu}{\chi^2/\nu}

\shorttitle{Gamma-Ray Variability of Cygnus X-1} \shortauthors{McConnell et al.}

\begin{document} \title{The Soft Gamma-Ray Spectral Variability of Cygnus X-1}

\author{M. L. McConnell\altaffilmark{1,9},
A. A. Zdziarski\altaffilmark{2},
K. Bennett\altaffilmark{5},
H. Bloemen\altaffilmark{4},
W.  Collmar\altaffilmark{3},
W.  Hermsen\altaffilmark{4},
L. Kuiper\altaffilmark{4},
W. Paciesas\altaffilmark{7},
B. F. Phlips\altaffilmark{6},
J. Poutanen\altaffilmark{8},
J. M. Ryan\altaffilmark{1},
V. Sch\"onfelder\altaffilmark{3},
H.  Steinle\altaffilmark{3},
and 
A. W. Strong\altaffilmark{3}}

\altaffiltext{1}{Space Science Center,University of New Hampshire, Durham, NH 03824}
\altaffiltext{2}{N. Copernicus Astronomical Center, Warsaw, Poland}
\altaffiltext{3}{Max Planck Institute for Extraterrestrial Physics, Garching, Germany}
\altaffiltext{4}{Space Research Organization of the Netherlands (SRON), Utrecht,The Netherlands}
\altaffiltext{5}{ Astrophysics Division, ESTEC, Noordwijk, The Netherlands} 
\altaffiltext{6}{Naval Research Laboratory, Code 7650, Washington, DC 20375}
\altaffiltext{7}{University of Alabama at Huntsville, Huntsville, AL  35899}
\altaffiltext{8}{Astronomy Division, P.O. Box 3000, 90014 University of Oulu, Finland}
\altaffiltext{9}{e-mail: Mark.McConnell@unh.edu}

\begin{abstract}

We have used
observations of Cygnus X-1 from the Compton Gamma-Ray Observatory (\gro\/) and 
BeppoSAX to study the variation
in the MeV $\gamma$-ray emission between the hard and soft spectral states, using
spectra that cover the energy range 20 keV up to 10 MeV. These data provide
evidence for significant spectral variability at energies above 1 MeV. In
particular, whereas the hard X-ray flux {\it decreases} during the soft
state, the flux at energies above 1 MeV {\it increases}, resulting in a
significantly harder $\gamma$-ray spectrum at energies above 1 MeV. 
This behavior is consistent with the
general picture of galactic black hole candidates having two distinct spectral
forms at soft $\gamma$-ray energies. These data extend this picture, for the
first time, to energies above 1 MeV. 
We have used two different hybrid thermal/non-thermal Comptonization models 
to fit broad band spectral data obtained in both the hard and soft spectral states.
These fits provide a quantitative estimate of the
electron distribution and allow us to probe the physical changes that take place
during transitions between the low and high X-ray states. We find that there is a
significant increase (by a factor of $\sim 4$) in the bolometric luminosity as the 
source moves from the hard state to the soft state.  Furthermore, the presence 
of a non-thermal tail in the Comptonizing electron distribution provides significant 
constraints on the magnetic field in the source region.

\end{abstract}

\keywords{accretion, accretion disks --- black hole physics --- gamma rays: observations --- stars: individual
          (Cygnus X-1) --- X-rays: stars}


\section{Introduction}

High energy emission from galactic black hole candidates (GBHCs) is characterized
by variability on time scales ranging from msec to months. In the case of Cygnus
X-1, it has long been recognized that, on time scales of several weeks, the soft
X-ray emission ($\sim 10$ keV) generally varies between two discrete levels
\citep[e.g.,][]{priedhorsky1983,ling1983,liang1983}. The source seems to spend
most ($\sim 90\%$) of its time in the so-called {\em low X-ray state},
characterized by a relatively low flux of soft X-rays and a relatively high flux
of hard X-rays ($\sim100$ keV).  This state is sometimes referred to as the {\em
hard state}, based on the nature of its soft X-ray spectrum.  On occasion, it
moves into the so-called {\em high X-ray state}, characterized by a relatively
high soft X-ray flux and a relatively low hard X-ray flux. This state is
sometimes referred to as the {\em soft state}, based on the nature of its soft
X-ray spectrum.  There are, however, some exceptions to this general behavior. 
For example, HEAO-3 observed, in 1979,  a relatively low hard X-ray flux
coexisting with a low level of soft X-ray flux \citep{ling1983,ling1987}. 
\citet{ubertini1991} observed a similar behavior in 1987.

Observations by the BATSE, OSSE, COMPTEL and EGRET 
instruments on the Compton Gamma-Ray Observatory (\gro\/),
coupled with observations by other high-energy experiments (e.g., SIGMA, ASCA and
RXTE) have provided a wealth of new information regarding the emission properties
of galactic black hole candidates. One important aspect of these high energy
radiations is spectral variability, observations of which can provide constraints
on models which seek to describe the global emission processes.  Based on
observations by OSSE of seven transient galactic black hole candidates at soft
$\gamma$-ray energies (i.e., below 1 MeV), two $\gamma$-ray spectral shapes have
been identified that appear to be well-correlated with the soft X-ray state
\citep{grove1997,grove1998,grove1999}. In particular, these observations define a
{\it breaking} $\gamma$-ray spectrum that corresponds to the hard (low) X-ray state and
a {\it power-law} $\gamma$-ray spectrum that corresponds to the soft (high) X-ray state.

A thorough understanding of the nature of these systems requires modeling
that cannot only explain the individual spectra, but can also explain the
transitions between the various spectral states \citep[e.g.,][]{grove1998,liang1998,poutanen1998a}.
In recent years, a general theoretical picture of the accretion flow in Cygnus
X-1 has emerged which appears to provide a reasonable explanation of the spectral
data in both the low and high X-ray states.  This model includes an inner
optically-thin, geometrically-thick advection-dominated accretion flow (ADAF)
surrounded by an outer, geometrically-thin, optically-thick accretion disk
\citep{esin1998}. The outer disk is characterized by a blackbody spectrum.  ADAF
flows \citep[e.g.,][]{narayan1996} are characterized by their relatively low
radiative efficiencies and by a two-temperature structure, with the ions nearly
virial at $T_i \sim10^{12}$ K and the electrons at $T_e \sim 10^9$ K.  The high
temperature of the ADAF leads to an extended, quasi-spherical geometry.  Hot
optically-thin ADAFs exist only below a certain critical accretion rate.  The
transition radius between the ADAF and the thin disk therefore depends on the
accretion rate.  At higher accretion rates, where it is more difficult to support
the ADAF, the transition radius moves to smaller radii, closer to the black hole.
The ADAF region is largely responsible for the hard X-ray flux ($\sim20-100$
keV), while the outer thin disk is generally responsible for the soft X-ray flux
($\sim2-10$ keV).

In the context of this general model, the spectral state of Cygnus X-1 depends on
the accretion rate.  At low accretion rates, the inner ADAF extends out to a
transition radius of $\sim100$ Schwarzschild radii \citep{esin1998}.  In this
configuration, the ADAF region makes a significant contribution to the hard X-ray
flux.  At higher accretion rates, it becomes more difficult to support the ADAF.
The ADAF region therefore shrinks, and the transition radius moves inward,
although there may exist a low-density ADAF corona surrounding the thin disk
\citep{narayan1998}.  The level of hard X-ray flux decreases due to the smaller
volume of the ADAF region, while the level of soft X-ray flux increases due to
the larger size of the thin disk region.  In this scenario, the hard state
corresponds to a relative low accretion rate, with the spectrum dominated by the
ADAF region, and the soft state corresponds to a relatively high accretion
rate, with the spectrum dominated by the blackbody of the outer thin disk region.
 Small changes in the accretion rate (on the order of 10--15\%) may be sufficient
to trigger a transition between the hard and soft states \citep{esin1998}.

The ADAF model described above provides a consistent framework for understanding
the essential dynamics and spectra of black hole accretion flows.  In the context
of this framework, however, simple thermal Comptonization models appear unable to
account for all of the spectral features, especially the hard power-law tail that
is seen at energies above $\sim600$ keV \citep{gierlinski1999}.  
\citet{poutanencoppi1998} used a geometry similar to that
described above \citep{poutanen1997} and assumed some (unspecified) source of
non-thermal electrons that remains constant during the spectral state
transitions.  This suggests that the non-thermal component may play a more
significant role, especially at higher energies, during the high X-ray state,
where the ADAF contribution is suppressed.

Hybrid
thermal/non-thermal plasmas have often been successfully used to model the
observed data \citep[e.g.,][]{gierlinski1999,poutanencoppi1998}.  
Based on the assumption that the
spectrum results from inverse Compton scattering of a
thermal photon spectrum by energetic electrons, the underlying electron population could be described as
a combination of a thermal Maxwellian and a power-law  tail extending to higher
energies.  
The presence of a
non-thermal component is often assumed a priori, without any specific model to
explain the origin, although the existence of such distributions is clearly 
established in the case of solar
flares \citep[e.g.,][]{coppi1999} and it is therefore natural to expect that
similar distributions exist elsewhere in the universe
\citep[e.g.,][]{crider1997,gierlinski1997,poutanen1996,poutanen1998a,poutanencoppi1998,coppi1999}.
Others have considered physical mechanisms
by which non-thermal electron distributions might be developed.  
For example, both stochastic particle acceleration \citep{dermer1996,li1996} and MHD turbulence 
\citep{li1997} have been proposed as mechanisms for directly accelerating the 
electrons.  The ion population might also contribute to the non-thermal 
electron distribution in the case where a two-temperature plasma develops
\citep[e.g.,][]{dahlbacka1974,shapiro1976,chakrabarti1995}.  
With ion population 
temperatures approaching $kT_i \sim 10^{12}$ K, $\pi^o$ production
from proton-proton interactions may take place \citep[e.g.,][]{eilik1980,eilik1983,mahadevan1997}.
The $\pi^o$ component may then lead, via photon-photon
interactions between the $\pi^o$-decay photons and the X-ray photons,
to production of energetic (nonthermal) $e^{+}-e^{-}$ pairs. 
\citet{jourdain1994} used this concept to 
fit the hard X-ray tails of not only Cygnus X-1, but also
GRO J0422+32 and GX 339-4, as measured by both SIGMA and OSSE. 
While retaining a  standard thermal Comptonization spectrum \citep{sunyaev1980}
to explain the emission at energies below
200 keV, they used  $\pi^o$ production to generate the nonthermal 
pairs needed to fit the spectrum at energies above $\sim 200$ keV.

The power-law spectra seen in the high X-ray state have also been modeled as
resulting from bulk-motion Comptonization
\citep[e.g.,][]{ebisawa1996b,titarchuk1997,laurent1999}. In this model, the flow
becomes quasi-spherical within the inner-most stable orbit. The nearly
relativistic flow of the free-falling electrons gives rise to the Comptonization
of ambient photons.  This model predicts power-law spectra, with a slope that
depends on the mass acretion rate.  The difficulty with this model is that it
predicts spectral sharp cutoffs below 500 keV, a result that is clearly
inconsistent with the observed spectra.  Although we cannot rule out bulk motion
Comptonization as a contributor to the spectrum at lower energies, it is clearly 
not capable of accounting for the high energy emission.

Improvements in the theoretical modeling of spectral state transitions can be
expected to arise from improved observations at energies above 600 keV.  It will
be important to understand how this part of the spectrum, most likely dominated
by non-thermal emission, changes during the spectral transition.  Of particular
interest will be observations that can discern a clear cutoff in the spectra at 
high energies.  
The precise energy of the cutoff is a function of the compactness of the source 
region, since it is influenced by $\gamma-\gamma$ opacity.  
A measure of the cutoff energy, possibly coupled
with measurements of the 511 keV $e^{\pm}$ annihilation line, will help 
constrain the compactness of the region responsible for the emission and determine 
the extent to which $e^{\pm}$ pairs may play a role in the emission region
\citep{poutanen1998a}.

Using hard state data collected during the first three years of the \gro\/ mission,
\citet{mcconnell2000a} compiled a broad-band  hard state spectrum of Cygnus X-1
using contemporaneous data from all four instruments on \gro\/ (BATSE, OSSE,
COMPTEL and EGRET).  Unlike previous broad-band studies, these data provided a
measurement of the spectrum at energies above 1 MeV.  The resulting spectrum
showed evidence for significant levels of non-thermal emission at energies out to
5 MeV.  The spectral shape, although consistent with the so-called breaking
spectral state \citep{grove1997,grove1998} of the $\gamma$-ray emission,
was clearly not consistent with
standard Comptonization models. The hybrid thermal/non-thermal model of
\citet{poutanen1996} was used to fit the  hard state data, with fits that indicated a thermal electron
population with a temperature of $\sim90$ keV and a high energy power-law electron component 
with a spectral index of $\sim4.5$.

In May of 1996, a transition of Cyg X-1 into a soft state was observed by
RXTE, beginning on May 10 \citep{Cui1997}.  The 2--12 keV flux reached a level of
2 Crab on May 19, four times higher than its normal value. Meanwhile, at hard
X-ray energies (20-200 keV), BATSE measured a significant {\it decrease} in flux
\citep{zhang1997}.  Motivated by these dramatic changes, a target-of-opportunity
(ToO) for \gro\/, with observations by OSSE and COMPTEL began on June 14
(\gro\/ viewing period 522.5). Here we report on the results from an analysis of
the \gro\/ data from this ToO observation, incorporating the high energy results 
from COMPTEL.  This includes a comparison with results obtained
from an updated analysis of \gro\/ soft state data, 
making use of the same data studied previously by \citet{mcconnell2000a}.
In $\S$ 2 we describe the \gro\/ observations of Cygnus X-1 in its hard state.  
The data analysis is described in $\S$ 3,
followed by a discussion of those results in $\S$ 4.

\section{Observations}

During its nine-year lifetime (1991--2000), the instruments on \gro\/ 
obtained numerous observations of the Cygnus region.
The COMPTEL experiment \citep{schoenfelder1993}, imaging the energy range from about 750 keV up to 30 MeV,
collected an extensive set of data, in part due to its rather large ($\sim \pi$
steradian) field-of-view.  The COMPTEL data currently provide the best available
source of data for studies of Cygnus X-1 at energies above 1 MeV.

The 20--100 keV time history of Cygnus X-1, as derived from BATSE
occultation data, is shown in the center panel of Figure 1. The top
panel of Figure 1 shows the 20--100 keV power-law spectral index, as
derived from the BATSE occultation data.  These data cover most of the
\gro\/ mission, from the launch in April of 1991 until the end of 1999. 
During the first few months of the \gro\/ mission (up
until October of 1991), all-sky monitoring data from Ginga (1--20 keV)
was available, showing that the source was in its low X-ray state during
this period \citep{kitamoto2000}.  From October of 1991 until December
of 1995, there were only sporadic pointed X-ray observations of the soft
X-ray flux from Cygnus X-1.  It was not until the launch of RXTE, in
December of 1995, that continuous data on the soft X-ray flux once again
became available.  The data from the RXTE All-Sky Monitor (ASM) are
shown in the lower panel of Figure 1, in the form of the 2--10 keV count
rate.

The data shown in Figure 1 dramatically demonstrate the general X-ray
behavior of Cygnus X-1.  During the \gro\/ mission, Cygnus X-1 spent 
about 90\% of its time in the hard state.  In
this state, the soft X-ray flux (2--10 keV) is relatively low, while the
hard X-ray flux (20--100 keV) is relatively high.  The spectral shape in
the 20--100 keV energy band is a relatively hard power-law spectrum with
a photon spectral index, $\Gamma$, near $1.8$.
The soft state was clearly observed during the \gro\/ mission on only two occassions.
In each case, the soft state period lasted about 5 months.  The soft state is
characterized by a  relatively high level of soft X-rays (2--10 keV), a
relatively low level of hard X-rays (20--100 keV) and a relatively soft spectrum
in the 20--100 keV energy band (photon spectral index $\sim 2.5$). The soft
state was first observed by \gro\/ in January of 1994, at a time (prior to
the launch of RXTE) when there was no soft X-ray monitoring data available.  
(This transition is clearly seen in Figure 1 near TJD 9400.) A
\gro\/ target-of-opportunity was declared (\gro\/ viewing period 318.1) so that all
four \gro\/ instruments (not just BATSE) could collect data.  Observations by
COMPTEL showed no detectable level of emission.  This null result, however, was
consistent with an extrapolation of the $E^{-2.7}$ power-law spectrum measured at
hard X-ray energies by both BATSE \citep{ling1997} and OSSE \citep{phlips1996}.

The second observation of a soft state took place in May of 1996.  The
transition was first observed by RXTE, beginning on May 10 \citep{Cui1997}.  The
2--12 keV flux reached a level of 2 Crab on May 19, four times higher than its
normal value. Meanwhile, at hard X-ray energies (20-200 keV), BATSE measured a
significant {\it decrease} in flux \citep{zhang1997}.  Motivated by these dramatic
changes, a ToO for \gro\/ was declared and observations by OSSE and COMPTEL began
on June 14 (\gro\/ viewing period 522.5). (Unfortunately, the EGRET experiment was
turned off during this viewing period, as part of an effort to conserve its
supply of spark chamber gas.) During the ToO, COMPTEL collected 11 days of data
(from June 14 to June 25) at a favorable aspect angle of 5.3$^{\circ}$. The X-ray
flux time histories near the time interval associated with VP 522.5 are shown in
Figure 2.

An early preliminary analysis of COMPTEL data from this second high-state
observation revealed some unusual characteristics \citep{mcconnell2000b}.  The 1--3 MeV image (Figure 3)
showed an unusually strong signal from Cygnus X-1 when compared with other
observations of similar exposure.  The flux level was significantly higher than
the average flux seen from earlier observations \citep{mcconnell1994,mcconnell2000a}.  In the
1--3 MeV energy band, the flux had increased by a factor of 2.5, from $8.6
(\pm2.7) \times 10^{-5}$ cm$^{-2}$s$^{-1}$ MeV$^{-1}$ to $2.2 (\pm0.4) \times
10^{-4}$ cm$^{-2}$ s$^{-1}$ MeV$^{-1}$.  The observed change in flux is
significant at a level of $2.6\sigma$.  In addition, unlike in previous
measurements, there was no evidence for any emission at energies {\it below} 1
MeV.  This fact is explained, in part, by a slowly degrading sensitivity of
COMPTEL at energies below 1 MeV due to increasing energy thresholds in the lower
(D2) detection plane.  Part of the explanation, however, appears to be the
presence of a much harder source spectrum.  OSSE data collected during this
period showed a photon spectrum similar to that observed in 1994 (i.e., a power-law with
an index of $\sim 2.5$), but at a higher intensity level, about a factor of two
higher in overall normalization.  The extrapolation of this more intense
power-law spectrum is entirely consistent with the positive detection by COMPTEL.

\section{Data Analysis}

A more detailed description of the COMPTEL data analysis is
given in \citet{mcconnell2000a}.  Here we provide only a brief overview.
The COMPTEL image shown in Figure 3 is a maximum
likelihood map derived from VP 522.5 data integrated over the energy loss range
of 1--3 MeV. The contours represent constant values of the quantity $-2
\ln{\lambda}$, where $\lambda$ is the likelihood ratio. In a search for single
point sources, $-2 \ln{\lambda}$ has a chi-square distribution with 3 degrees of
freedom.  (For instance, a $3\sigma$ detection corresponds to $-2 \ln{\lambda}$ =
14.2.)  Cygnus X-1 is clearly visible.  The likelihood reaches a value of $-2
\ln{\lambda} = 30.1$ at the position of Cygnus X-1, which corresponds to a
detection significance of $5.5\sigma$. These same data were used to derive the
$1\sigma$, $2\sigma$ and $3\sigma$ location confidence contours shown in Figure
4, which demonstrate the ability of COMPTEL to locate the source of emission.  In
defining constraints on the source location, $-2 \ln{\lambda}$ has a chi-square
distribution with 2 degrees of freedom.  So the $1\sigma$, $2\sigma$ and
$3\sigma$ location confidence contours correspond to a change in likelihood of
2.3, 6.2, and 11.8, respectively. The contours reflect only the statistical
uncertainties; systematic effects are not included.  The COMPTEL flux results 
for VP 522.5 are shown in Table 1.

The analysis of COMPTEL data for a weak source (such as Cygnus X-1) involves
generating an image for each of several energy bands, deriving the source flux in
each energy band, and subsequently combining these results into a spectrum
\citep[e.g.,][]{mcconnell2000a}.  The image generation process, in turn, requires
an instrument point-spread-function (PSF) that is dependent on some assumed form
for the incident photon spectrum.  Because the spectrum extraction relies on an
assumed source spectrum (we typically assume an $E^{-2}$ power-law spectrum), it
is not possible to analyze the COMPTEL data using a simple response function to
relate measured energy-loss count rates to the incident photon flux. We have
therefore resorted to spectral fitting of COMPTEL data in photon space.  We have
previously shown \citep{mcconnell2000a} that this approach to COMPTEL spectral
analysis works fine for the range of spectra considered for Cygnus X-1. In other
words, for the range of parameters considered here, there is no evidence of any
significant level of spectral compliance in the COMPTEL spectral analysis.

In previous work \citep{mcconnell2000a}, we analyzed a contemporaneous set of \gro\/
data corresponding to the hard state of Cygnus X-1.  That analysis was performed
entirely in photon space, using one deconvolved spectrum each for BATSE, OSSE and
COMPTEL.  The BATSE spectrum had been generated using the JPL Enhanced BATSE
Occultation Package \citep[EBOP;][]{ling1996, ling2000}, while the OSSE spectrum was
based on the results of \citet{phlips1996}. The analysis was performed within
XSPEC (Arnaud 1996) to take advantage of the XSPEC analysis tools.  In this case, however, the
spectral data ({\tt pha} files) were generated in units of photons cm$^{-2}$ s$^{-1}$
and the response function matrices ({\tt rsp} files) were generated as unit matrices. 
With these data, the spectral fits were effectively being performed in photon
space.  This approach greatly simplified the analysis effort.

The analysis employed here for the soft state data from VP 522.5
represents a significant improvement over that performed previously in generating
a broad-band gamma-ray spectrum for the hard state.  Although the
limitations of the COMPTEL data analysis remain, we have utilized more complete
spectral response information for both BATSE and OSSE, using proper XSPEC {\tt pha} and
{\tt rsp} files.  The fundamental nature of the COMPTEL data (in particular, its
reliance on an assumed PSF for extracting source counts) still precludes a 
proper XSPEC analysis of the COMPTEL data.

Since the BATSE EBOP processing has not been carried out for data collected after
1994, EBOP data for VP 522.5 is not available.  Instead, we have used data
derived from the BATSE team's standard Earth occultation analysis
\citep{harmon2002}.  The final BATSE spectrum represents a weighted average of the 
four forward-facing detectors.

For consistency, in order to make a more useful comparison with the
soft state data, we have repeated the earlier hard state analysis \citep{mcconnell2000a}
following the
same procedures as we have used here for the VP 522.5 data. In particular, 
the updated hard state
analysis now also used BATSE data derived from the standard Earth occultation technique
\citep{harmon2002}.  Data from nine separate \gro\/ viewing periods were used 
\citep[see Table 1 in][]{mcconnell2000a}. 
In each case a weighted average spectrum was derived from the data for use in the final 
analysis.

Finally, our most recent analysis incorporates the BATSE data down to 20 keV 
and OSSE data down to 50
keV. Previously, we had used only those data above 200 keV. The lower energy
threshold of this analysis improves the ability of our fits to constrain the
spectral models.  At the same time, the lower threshold may also make the 
analysis more sensitive to systematic uncertainties in the low energy response 
of both BATSE and OSSE.  The OSSE (Johnson et al.\ 1993) data include energy-dependent 
systematic errors
(estimated from the uncertainties in the low-energy calibration and response of
the detectors using both in-orbit and prelaunch calibration data), which are
most important at the lowest energies, $\sim 3\%$ at 50 keV, decreasing to
$\sim 0.3\%$ at $\ga 150$ keV. To the BATSE data, we added a 5\% systematic
error. The COMPTEL data have relatively large statistical errors, and thus no
systematic error was added.

\subsection{The Average Hard (Low) State Spectrum}
\label{s:hard}

X-rays from Cygnus X-1 in the hard state are well modeled by thermal
Comptonization and Compton reflection (Gierli\'nski et al.\ 1997; Di Salvo et
al.\ 2001; Frontera et al.\ 2001). Thus, we fit
the joint data by this model, but also allowing for a tail at high
electron energies rather than the Maxwellian cutoff.  This is similar to the 
approach used in \citet{mcconnell2000a}, except that here we have added the 
Compton reflection component, which is important at energies below a 
few hundred keV.

We first fit the data using the Comptonization
model ({\tt compps})\footnote[1]{available at
ftp://ftp.astro.su.se/pub/juri/XSPEC/COMPPS} 
of Poutanen \& Svensson (1996), assuming a spherical source
geometry. The electrons have the total Thomson optical depth of $\tau$. Their
distribution in this model is Maxwellian with an electron temperature, $kT$, up
to a Lorentz factor, $\gmin$, above which it is a power law with an index, $p$.
The power law extends to a large Lorentz factor, $\gamma_{\rm max}$. 
The precise value of $\gamma_{\rm max}$,
however, has little effect on the fit to our data as long as $\gamma_{\rm
max}^2$ times the seed photon energy is $> 10$ MeV. Given that the seed photons
peak at a fraction of keV \citep{ebisawa1996a,disalvo2001}, we assume $\gamma_{\rm
max}=10^3$. The Comptonization spectrum is then Compton-reflected from a cold
slab (presumably an accretion disk) subtending a solid angle, $\Omega$
(Magdziarz \& Zdziarski 1995). A disk inclination of $i=45\degr$ is assumed
\citep[as in][]{gierlinski1997,frontera2001}.  For this model, we found $\Omega$ is
not constrained by our data and we kept it fixed at a
typical value $\Omega/2\pi = 0.5$ \citep{gierlinski1997,gilfanov1999,disalvo2001}.
This value also follows from our fit below
using another theoretical model. The seed photons for Comptonization are
assumed to be a (multicolor) blackbody emission of the disk with the maximum
blackbody temperature of $kT_{\rm s}=0.2$ keV, which approximately corresponds
to a single blackbody temperature of $\sim 0.13$--0.15 keV obtained in the fits of
\citet{ebisawa1996a} and \citet{disalvo2001}.

During the data analysis, we find some residual discrepancies between the
different data sets, as expected for different instruments. First, the BATSE
spectrum has a slightly higher normalization than the OSSE one, and thus we
allow their relative normalization to be free in the fits. Furthermore, the
BATSE spectrum is systematically slightly softer, by $\Delta\Gamma\simeq 0.1$
(where $\Gamma$ is the photon power law index), than the OSSE spectrum. We find that
multiplying  the BATSE model by an additional power law with that
$\Delta\Gamma$ leads to a reduction of $\cnu$ from 71/48 to 37/47, which is
highly significant at the $2\times 10^{-8}$ level, using the F-test (Bevington
\& Robinson 1992). Thus, we apply this correction, fixing $\Delta\Gamma$
hereafter at the best-fit value. The best-fit ratio of the BATSE and OSSE
fluxes at 100 keV is then 1.26 (but higher and lower at higher energies and
lower energies, respectively). Furthermore, we find that the COMPTEL data at $\sim
1$ MeV appear to have somewhat higher normalization than the BATSE and OSSE
ones. The best-fit relative normalization is $\sim 1.5$, consistent with
results for the soft state (see below). Given the limited statistics of the
COMPTEL hard-state data, we fix that relative normalization at 1.5. We note
that this yields a conservative estimate of the amplitude of the nonthermal
tail in Cygnus X-1.

The {\tt compps} fit results are given in Table \ref{t:pars}. They are similar to
the preliminary results of \citet{mcconnell2000a}, who neglected Compton reflection. We note that
the fitted electron distribution is allowed to be significantly different from
a pure Maxwellian, with a power law tail beginning at a rather low energy. This
reflects the fact that an arbitrary electron distribution peaked at low
energies yields Comptonization spectra relatively similar to those from a pure
Maxwellian \citep{ghisellini1993}, apart from the high-energy tail
in the former case.

We then fit the same data using a different hybrid Comptonization model, {\tt
eqpair} \citep{coppi1992,coppi1999,poutanencoppi1998,gierlinski1999}. Unlike {\tt compps},
which assumed the form of the steady-state electron distribution, {\tt eqpair}
calculates that distribution self-consistently assuming instead the electron
acceleration to be a power law with an index, $\Gamma_{\rm inj}$ between
$\gmin$ and $\gamma_{\rm max}$. The accelaration takes place  in a background
thermal plasma with a Thomson optical depth of ionization electrons, $\tau_{\rm
i}$. The steady-state electron distribution consists then of a Maxwellian at
the temperature, $kT$, calculated from the balance of Compton and Coulomb gains
and losses, and the optical depth, $\tau$, with $\tau - \tau_{\rm i}$ due to
the \ee\ pair production. The nonthermal steady-state electron distribution is
calculated from Coulomb and Compton losses of both the accelerated electrons
and \ee\ pairs produced at nonthermal energies. That distribution, in general,
does not have a power law form. Unlike {\tt compps}, the electron distribution
is now a sum of the Maxwellian and the nonthermal part (rather than being a
Maxwellian up to $\gmin$ and then nonthermal). We assume $\gmin=1.5$ and
$\gamma_{\rm max}=10^3$. Unlike the case of {\tt compps}, where the value of
$\gmin$ determined the transition from the Maxwellian to the power law, that
value has relative little effect on the fit now.

The rates of microscopic processes per unit light travel time across the source
depend in general on the plasma compactness, $\ell\equiv {\cal  L}\sigma_{\rm
T}/({\cal R} m_{\rm e} c^3)$, where ${\cal L}$ is a power  supplied to the hot
plasma, ${\cal R}$ is its characteristic size, and  $\sigma_{\rm T}$ is the
Thomson cross section (e.g., Svensson 1987). We then define a hard compactness,
$\lh$, corresponding to the power supplied to the electrons, and a soft
compactness, $\ls$, corresponding to the power in soft seed photons irradiating
the plasma (which are assumed to be emitted by a blackbody disk). The
compactnesses corresponding  to the electron acceleration and to a direct
plasma heating (i.e., in addition to Coulomb energy exchange with nonthermal
\ee\ and Compton heating) of the  thermal \ee\ are denoted as $\lnth$ and
$\lth$, respectively, and $\lh=\lnth +  \lth$. Details of the model are given
in \citet{gierlinski1999}.

The {\tt eqpair} fit results are given in Table \ref{t:pars}. Figure \ref{f:hard} shows
the spectrum. Figure \ref{f:hard_m} shows the model spectral components over the broad 
energy range from 
0.1 keV to 100 MeV. The nonthermal high-energy tail starts at $\sim
1$ MeV.

The bolometric flux derived from the {\tt eqpair} model 
is very similar to that in the {\tt compps} model. Some
differences between the values of $kT$ and $\tau$ may be  attributed to the
different treatment of the microphysics (see above). Also, the {\tt eqpair} model gives a
somewhat better fit to the data than the {\tt compps} model.  In addition, 
the {\tt eqpair} model provides a better constraint on the value of $\Omega/2\pi$ 
(Table \ref{t:pars}). The better fit most likely reflects the fact that
more physical processes are accounted for by {\tt eqpair} than by {\tt
compps}.

In particular, pair production is important at the best fit of the {\tt eqpair}
model, which accounts for $\tau- \tau_{\rm i} >0$ in Table \ref{t:pars}, and
the associated injection of nonthermal \ee\ at low energies. The latter effect
leads to a softening of the nonthermal spectra (Svensson 1987) and explains the
relatively low value of $\Gamma_{\rm inj}$ (without pair production and in the
Thomson regime, $p$ would be $\simeq \Gamma_{\rm inj}+1$, e.g., Blumenthal \&
Gould 1970). On the other hand, at the lowest $\ls$ allowed by the data,
$\simeq 0.2$, the plasma compactness is so small that we find basically no pair
production, i.e., $\tau= \tau_{\rm i}$. Thus, the present data do not resolve
the issue of the role of pair production conclusively.

The bolometric luminosity of the average hard-state spectrum, $L$, equals
$\sim 1\%$ of the Eddington luminosity, $L_{\rm E}\simeq 1.5(M/{\rm M}_\odot)
\times 10^{38}$ erg s$^{-1}$, assuming isotropy, a black-hole mass of $M= 10
{\rm M}_\odot$ and a distance of 2 kpc (see discussion and references in
Gierli\'nski et al.\ 1999). The best-fit total compactness,
$\lh+\ls\sim 30$, corresponds to the characteristic dimension of the plasma of
$\sim 10^2 GM/c^2$ under the assumptions as above. The X-ray spectrum is rather
hard, with the amplification of the seed photons by the factor $\lh/\ls\simeq
17$. Only a small fraction of the power supplied to the plasma,
$\lnth/\lh\simeq 0.08$, is used for nonthermal electron acceleration.

\subsection{Broad-Band Spectrum in the Soft (High) State}
\label{s:soft}

As in the case of the hard state, the broad-band spectrum of Cygnus X-1 in the soft
state is well fit by emission from a blackbody disk, Compton scattering by
thermal and nonthermal electron components, and Compton reflection with the
accompanying Fe K$\alpha$ fluorescence line \citep{gierlinski1999,frontera2001}. 
However, unlike the
hard state, the \gro\/ data alone ($\geq 20$ keV) cannot determine the
parameters of the thermal electron distribution. The reason for this is that,
whereas scattering by the thermal electrons dominates up to several hundred keV
in the hard state (see above, also Gierli\'nski et al.\ 1997), it dominates
only up to $\sim 10$ keV in the soft state \citep[][see below]{gierlinski1999}. Then, in the
\gro\/ energy range, the spectrum is {\it entirely\/} due to the emission of
the nonthermal electrons and Compton reflection.

Thus, in order to determine the parameters of the electron distribution
(including its thermal part) implied by the \gro\/ data in the soft state, we
combine them with the \sax\/ data from 1996 June 22 \citep{frontera2001}.
These data cover roughly a 90 minute time span during the much longer 11-day 
\gro\/ observation (June 14--25). For the \sax\/ observation, data from three
instruments, LECS, HPGSPC and PDS, are usable \citep{frontera2001}, extending the 
measured energy range down to 0.5 keV. We allow for a free
relative normalization of each set of spectral data with respect to that of 
OSSE. All the normalization factors are found to be $\sim 1$.

We use the same two models ({\tt compps} and {\tt eqpair}) as for the hard state. 
However, since our data
extend now down to $\sim 0.5$ keV, we let $kT_{\rm s}$ free. For the same
reason, we need to include the fluorescent Fe K$\alpha$ emission, present in
both states of Cygnus X-1. Since the line is produced by Compton reflection, we
need to relate its flux to the strength of Compton reflection. We follow here
results of George \& Fabian (1991) and \.Zycki \& Czerny (1994) and tie the
line flux to $\Omega/2\pi$ in such a way that the equivalent width with respect
to the total continuum is $\simeq 120$ eV when $\Omega/2\pi= 1$. Both the
line and the reflection continuum are assumed to come from an accretion disk
extending down to $6 GM/c^2$ \citep[e.g.,][]{gierlinski1999} and with the
reflection/fluorescence emissivity following that of a standard thin disk
(Shakura \& Sunyaev 1973). This results in a relativistic smearing (Fabian et
al.\ 1989) of both of those spectral components. The reflecting surface is
allowed to be ionized \citep{gierlinski1999,disalvo2001}, 
with the degree of ionization characterized
by the ionization parameter, $\xi\equiv 4 \pi F_{\rm ion}/n$ (where $F_{\rm
ion}$ is the ionizing flux and $n$ is the reflector density), and at the
temperature of $\sim kT_{\rm s}$. The elemental abundances are of Anders \&
Ebihara (1982).

Both {\tt compps} and {\tt eqpair} models provide very good descriptions of our
broad-band spectrum. Table \ref{t:pars} gives the fit results and Figure
\ref{f:soft} shows the spectrum for the {\tt eqpair} model. Figure \ref{f:soft_m}
shows the spectral components of the {\tt eqpair} fit to the spectrum. Both models predict
the power-law--like emission extending with no cutoff up to 10 MeV, in
agreement with the data.

Strong Compton reflection with $\Omega/2\pi \sim 1.3$ is seen, similar to the
results of \citet{frontera2001} and those from \xte\/ (Gilfanov et al.\ 1999). 
A likely cause of
$\Omega> 2\pi$ is relativistic anisotropy of Compton scattering 
\citep[see a discussion in][]{gierlinski1999}.
Scattering by nonthermal electrons, forming a power-law
like component, dominates a peaked component from thermal scattering at
energies above several keV \citep[as found by][]{gierlinski1999}.

Pair production is unimportant at the best fit. Thus, both models give
results fully consistent with each other. The values of $kT$ are virtually
identical, and the small difference in the values of $\tau$ is an artifact of
the different treatment of the radiative transfer, and $p\simeq \Gamma_{\rm
inj}+1$ (as expected for dominant Compton cooling in the Thomson regime, see
\S \ref{s:hard}).

The bolometric luminosity is about 4 times that in the hard state, and it is
$\sim 0.04 L_{\rm E}$ under the same assumptions as in \S \ref{s:hard}. 
In contrast to the hard state, only a small fraction of the total luminosity,
$\lh/(\lh+\ls)\simeq 0.15$, is emitted by the plasma outside the
optically-thick accretion disk, although part of the disk emission is due to
reprocessing of the hard, plasma, emission \citep[see discussion of the energy
balance in][]{gierlinski1999}. Also in contrast to the hard state, most, $\sim 0.7$, of the
power supplied to the plasma is used for nonthermal acceleration. Although the
electron temperature is very similar in both states, the optical depth in the
soft state is $\ll$ that in the hard state.

\section{Discussion}

The COMPTEL data alone can be used to draw some important conclusions regarding
the MeV variability of Cyg X-1.  Most importantly, the flux measured by COMPTEL
at energies above 1 MeV was observed to be {\em higher} (by a factor of 2.5) during the
soft state (in May of 1996) than it was during the hard state (as averged over several 
\gro\/ observations).  This 
is in contrast to the {\em lower} flux level observed at hard X-ray energies 
(i.e., near 100 keV) during the soft state.  The
lack of any detectable emission by COMPTEL below 1 MeV (i.e., in the 750 keV to 1 
MeV energy band) further suggests a hardening of the $\gamma$-ray spectrum during the 
soft state.

Inclusion of the BATSE and OSSE spectra adds considerably more information regarding the 
spectral variability.  
Whereas the low-state \gro\/ spectrum shows the breaking type spectrum
that is typical of most high-energy observations of Cyg X-1 \citep[e.g.,][]{mcconnell2000a}, 
the high-state \gro\/ spectrum shows the
power-law type spectrum that is characteristic of black hole candidates
in their high X-ray state.  
Our analysis of the soft state data from BATSE, OSSE and COMPTEL shows that the 
spectrum at these energies can be described by a single power-law with a best-fit 
photon spectral index of $\Gamma = 2.58 \pm 0.03$. 
A similar spectrum had already been reported 
for this same time period (VP 522.5) based on independent studies with data from 
both BATSE \citep{zhang1997} and OSSE \citep{gierlinski1997,gierlinski1999}. 
A detailed study of the broadband soft state spectrum, based on data from
ASCA, RXTE and \gro\//OSSE, was reported by \citet{gierlinski1999}, but they
did not include the higher energy COMPTEL data.  The inclusion of the
COMPTEL data in the high state spectrum provides evidence, for the first
time, of a continuous power-law (with a photon spectral index of 2.6)
extending beyond 1 MeV, up to $\sim10$ MeV.  No clear evidence for a cutoff
in the power-law spectrum can be discerned from these data.

A  power-law spectrum had also been observed by both OSSE and BATSE during the
high X-ray state of February, 1994 \citep[\gro\/ VP 318.1;
][]{phlips1996,ling1997}. These earlier data correspond to the low level of hard
X-ray flux near TJD 9400 in Figure 1. The spectrum observed during the 1994 high
state showed a similar photon spectral index ($\Gamma = 2.72$ vs. $\Gamma = 2.57$ for the 1996 high
state spectrum), but the overall intensity of the power law
was considerably lower \citep{gierlinski1999}.  Near 1 MeV, for example, the spectral amplitude was
about 3 times lower in 1994  than it was in 1996. This explains why Cygnus X-1 
was not observed by COMPTEL during the 1994 high
state.  The extrapolation of the lower-intensity
power-law fell below the sensitivity limit of COMPTEL. On the other hand,
the intensity observed in 1996 was sufficiently high to allow for a measurement of the spectrum
by COMPTEL.

We have used two different hybrid thermal/non-thermal Comptonization models 
({\tt compps} and {\tt eqpair}) to 
fit broad band spectral data obtained in both the hard and soft spectral states.
For the hard state analysis, we used data from \gro\/ covering 20 keV up to 10 MeV.
For the soft state analysis, we augmented the \gro\/ data with lower energy data from
BeppoSAX to provide improved constraints on the spectrum at energies down to 0.5 keV.
These fits provide a quantitative estimate of the
electron distribution and allow us to probe the physical changes that take place
during transitions between the low and high X-ray states. 
Hybrid Comptonization models have also been used to model the spectra
of other black hole binaries in their soft state, such as GRS 1915+105 \citep{aaz01}.

The high energy spectrum of Cygnus X-1 
 cannot be described by the bulk-motion Comptonization model
alone, which predicts a sharp cutoff above $\sim 100$ keV (Laurent \& Titarchuk
1999).  The hybrid comptonization models provide an adequate fit to the 
data without requiring any contribution from
bulk-motion Comptonization. 
Furthermore, the bulk-motion Comptonization power-law for $L\sim 0.04 L_{\rm E}$
corresponding to the soft state of Cygnus X-1 (see below), was found by Laurent
\& Titarchuk (1999) to be very soft, with $\Gamma\simeq 3.5$ at , i.e., much
softer than the observed $\Gamma\simeq 2.5$ \citep{gierlinski1999,frontera2001}.  
(See also the
discussion in Zdziarski 2000.) Note that the XSPEC model of bulk-motion
Comptonization, {\tt bmc} (Shrader \& Titarchuk 1998), does not include any
high-energy cutoff and thus cannot be applied to our data (or any data
extending to $\ga 100$ keV).

Figure \ref{f:compare} shows a comparison of the spectra in the two states. For
the hard state, we also show a typical spectrum at energies $\la 25$ keV
\citep[\sax\/ data from][]{disalvo2001}. 
We see that the two broad-band spectra cross each other
at $\sim 10$ keV and $\sim 1$ MeV. The dashed curve shows the model obtained by
fitting the hard-state data from \gro\/ only (\S \ref{s:hard}), and assuming
$\nh=6\times 10^{21}$ cm$^{-2}$. We see that this model predicts the low-energy
\sax\/ data relatively well, underestimating somewhat the observed spectrum
only at $\la 10$ keV due to the presence of a pronounced soft X-ray excess
present in the hard state \citep{ebisawa1996a,frontera2001,disalvo2001}, which is neglected
in our model fitted to the data at $\geq 20$ keV.

The bolometric flux or luminosity ratio between the soft state in June 1996 and the
average for the hard state is $\sim 4$.  This value is much more than the rough estimate of
$\sim 1.5-1.7$ based on the ASM and BATSE occultation results \citep{zhang1997}, but is consistent 
with the results of \citet{frontera2001}, based on studies with BeppoSAX. Such a
large value makes models of the state transition based on a change
of accretion rate plausible. Given the larger luminosity in the soft state, the
characteristic dimension of the hot plasma in the soft state based on the
compactness fit is similar to that in the hard state, $\sim 10^2 GM/c^2$.

These  data tend to support the general picture that the
transition between the hard and soft states  
results from a change in the disk transition radius between a hot inner corona (ADAF) and 
a cooler outer thin disk \citep[e.g.,][]{esin1998,narayan1998,poutanen1998a,poutanen1998b,poutanencoppi1998}.
In the hard state, this transition radius is relatively far from the black hole
(at $\sim 100$ Schwarzschild radii).  The spectrum is dominated by  
Comptonization off the thermal electrons in the hot inner corona.  Radio emission is
also more pronounced in this state \citep{fender2001}, 
with evidence for a radio emitting relativistic jet
\citep{stirling2001}.  As the transition 
radius moves inward, perhaps due to an increase in the accretion rate, the optically 
thick cool disk intercepts a larger fraction of the energy.  The thermal energy dissipation 
in the corona is reduced considerably and the blackbody disk component (the principal
component at soft X-ray energies) becomes more pronounced.

Although our data tend to support the above picture, we have not attempted to model the geometry in detail, since the precise geometrical configuration of the emitting region is largely unknown.  Furthermore, our new data cover the energy range near 1 MeV where geometry effects are difficult to study. One of the primary goals of the present paper is to determine the electron distribution of the radiating plasma. Our assumption of a spherical source geometry provides the necessary physics that is required to extract information on the electron spectrum.  We have further presumed that the thermal and nonthermal electrons are in the same physical region.  This assumption is based, in part, on the observations that show a negative correlation between the thermal and nonthermal components.  This need not be the case in reality, however, but the present data cannot be used to determine the extent to which the two populations are co-located.   A more detailed discussion of geometrical effects in the context of the \texttt{eqpair} model, including Compton reflection and energy balance, can be found in \citet{gierlinski1999}.

The shape of the electron distribution  and its high energy tail can
best be determined by measurements that extend into the MeV energy region.
The high-energy cutoff is related to the compactness of the source
region, since it depends, in part, on the influence of $\gamma-\gamma$ pair
production.  If $\gamma-\gamma$ pair production is an important source of opacity,
this would imply the presence of a significant level of $e^{\pm}$ pairs in the
source region.  In this way, a measure of the high-energy cutoff can
help determine the nature of the emitting plasma ($e-p$ or $e^{\pm}$).  Although
a measure of $e^{\pm}$ annihilation radiation can also serve as a diagnostic of
a pair plasma, it is likely that any annihilation radiation that may be
present would be considerably broadened (and perhaps blue-shifted), and
hence may not be readily observable.  Measurements to date with HEAO-3
\citep{ling1989} and with OSSE \citep{phlips1996} provide only
upper limits, or, at best, a marginal ($1.9\sigma$) detection 
\citep{ling1989} to the level of $e^{\pm}$ annihilation radiation. This further
underscores the need to define the high-energy cutoff as perhaps the
best means for constraining the source compactness and the nature of the
emitting region.  If INTEGRAL, with its improved line sensitivity,
succeeds in measuring an annihilation feature, then constraints on the
high-energy cutoff will be even more valuable.

The presence of a non-thermal tail in the electron distribution can also
provide constraints on the strength of the magnetic field in the source
region.  As pointed out by Wardzi\'nski \& Zdziarski (2001), the presence of even a weak
nonthermal electron tail increases strongly the emissivity of the
cyclo-synchrotron process with respect to the pure thermal case. If the
Compton-scattering electrons in Cygnus X-1 were purely thermal, that process
appears in general to be too inefficient to provide all of seed photons for the
Comptonization under simple assumptions of equipartition (Wardzi\'nski \&
Zdziarski 2000). Since we do see a blackbody component at low energies 
\citep{ebisawa1996a,disalvo2001}, this inefficiency is consistent with the seed photons for
Comptonization provided by the blackbody rather than by the cyclo-synchrotron
photons. On the other hand, the tail parameters obtained by \citet{mcconnell2000a}
yielded such a
copious supply of cyclo-synchrotron seed photons that the corresponding
luminosity would become $\sim 10^2$ times that observed (Wardzi\'nski \&
Zdziarski 2001). This conclusion is confirmed for the tail parameters fitted
here (G. Wardzi\'nski, private communication). Thus, either the magnetic field
in Cygnus X-1 is substantially below equipartition (at least an order of
magnitude) or the observed photon tail has a different origin than that due to
a high energy electrons.  In either case, this has important implications for 
models of the accretion flow in Cygnus X-1.

These studies also have implications that go beyond that of studying individual 
black hole sources.  Given the close spectral similarity between black-hole binaries in the hard
state and Seyferts (e.g.,\ Zdziarski 2000), it is possible that similar tails
are present in the spectra of the latter objects. 
\citet{stecker1999} have suggested that the hard tail emission 
seen in sources like Cygnus X-1 might account for an important component 
of the cosmic diffuse background radiation in the 200 keV -- 3 MeV energy band
\citep[see also][]{stecker2001}.
Note, however, that the tail
of Cygnus X-1 above 1 MeV contains relatively little flux, 1.3\% of the
bolometric (model) flux, for the fit with {\tt eqpair}. If a similar value is
characteristic of Seyferts, the combined emission from their high energy tails
may be too weak to account for the observed extragalactic MeV background,
perhaps arguing against the proposal by \citet{stecker1999}.

The next major satellite for this energy range, INTEGRAL, is expected to
have only slightly better continuum sensitivity than COMPTEL at energies near 1 MeV
with both its IBIS and SPI experiments 
\citep{schoenfelder2001}.  Furthermore, the much narrower FoV of the
INTEGRAL instruments ($\sim15\degr$) will mean that there will likely be only a
limited number of observations of Cygnus X-1.  This is in stark contrast
to the COMPTEL situation, in which the large FoV of COMPTEL ($\sim60\degr$)
resulted in many weeks of exposure, most of which were obtained during
the low X-ray state.  Given the large low-state exposure of \gro\/, 
it is quite likely that INTEGRAL may not be able to offer any
significant improvement in our knowledge of the hard state continuum 
spectrum at MeV energies. The
\gro\/ data may therefore provide the best view of the hard state MeV
continuum for many years to come.  
However,
COMPTEL is very limited in the data that it collected for the {\em soft state}
spectrum.  Additional  soft state observations with INTEGRAL could therefore
prove valuable. An important goal would be to search for a cutoff in the 
energy spectrum.  Pinning down the energy of this cutoff would be a
very important next step in our understanding of the high energy spectrum of
Cygnus X-1.  In this regard, INTEGRAL may be an extremely useful tool 
for collecting additional  soft state spectral data, providing that suitable 
target-of-opportunity observations can be acquired.

\acknowledgments

We thank P. Coppi and M. Gierli\'nski for their work on the {\tt eqpair} model.
The COMPTEL project is supported by NASA under
contract NAS5-26645, by the German government through DLR grant 50 Q 9096 8 and
by the Netherlands Organization for Scientific Research NWO.  This work has also
been supported at UNH by the \gro\/ Guest Investigator Program under NASA grant
NAG5-7745.  AAZ has been supported by grants from KBN (2P03C00619p1,2, 5P03D00821)  
and the Foundation for Polish Science. JP has been supported by the
Swedish Natural Science Research Council and the Anna-Greta and Holger
Crafoord Fund.

\clearpage
\begin{figure} 
\plotone{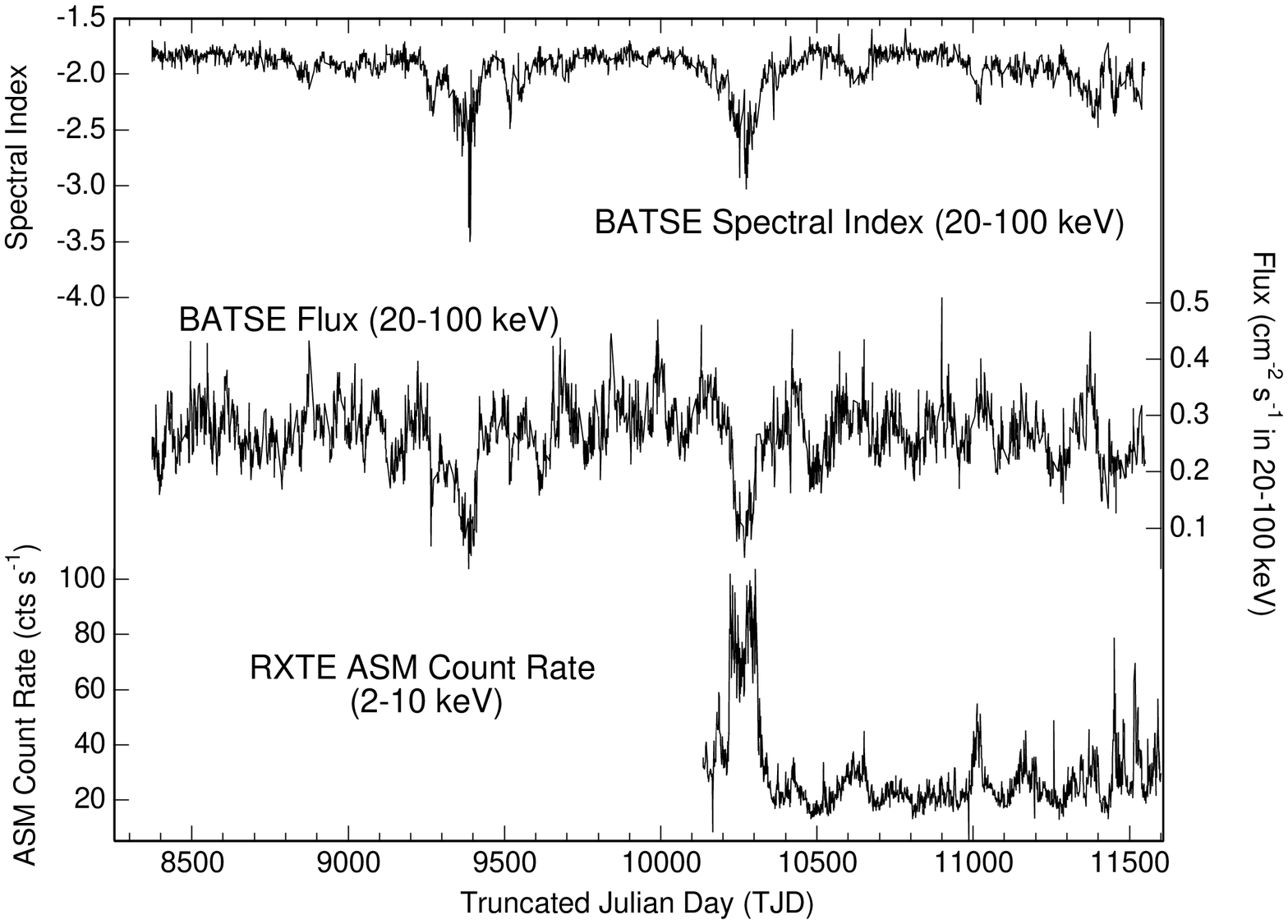} 
\caption{X-ray time histories
of Cygnus X-1 covering nearly the entire \gro\/ mission. The hard X-ray data come
from BATSE data that are derived from Earth occultation analysis in the 20--100
keV energy range.  The soft X-ray data (2--10 keV) are from the All-Sky Monitor
(ASM) on the Rossi X-Ray Timing Explorer (RXTE).} \end{figure}

\begin{figure} 
\plotone{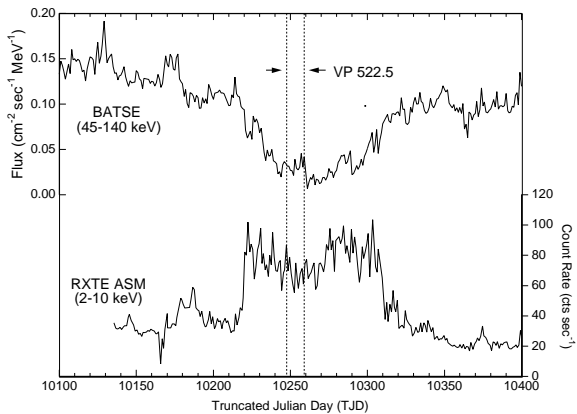} 
\caption{The time interval
of \gro\/ Viewing Period 522.5 is shown relative to the hard (upper) and soft 
(lower) X-ray time
histories from BATSE and the RXTE-ASM, respectively.  Note the very rapid
transition into and out of the high X-ray state as seen in soft X-rays.}
\end{figure}

\begin{figure} 
\plotone{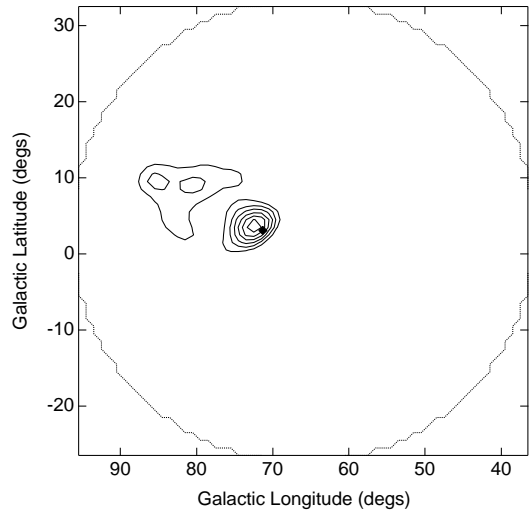} 
\caption{COMPTEL imaging of the
Cygnus region as derived from 1--3 MeV data collected during the soft state
of June, 1996 (\gro\/ viewing period 522.5). 
The outer contour (dotted line) represents the effective FoV of COMPTEL 
(with a 30$^{\circ}$ radius).  The remaining contours represent constant values of the quantity 
$-2\ln{\lambda}$, where $\lambda$ is the likelihood ratio. The contours start at a
value of 15, with a step size of 5.  The likelihood reaches a value of 30.1 at
the location of Cygnus X-1 (denoted by the diamond).} \end{figure}

\begin{figure} 
\plotone{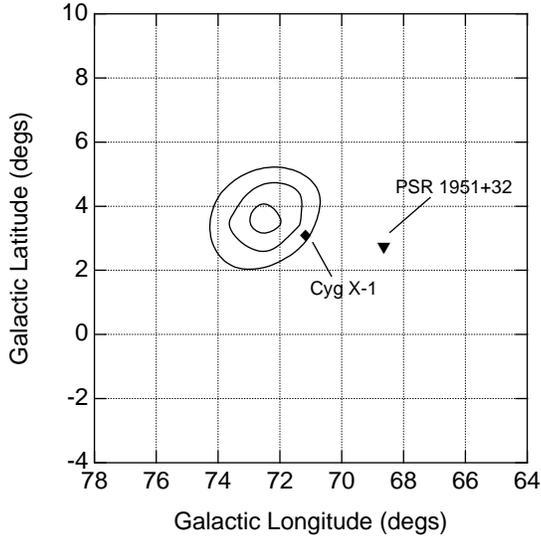} 
\caption{The 1, 2 and 3-$\sigma$
location contours derived from the likelihood map in Figure 3.  The emission is
consistent with a point source at the location of Cyg X-1, with no significant
contribution from PSR 1951+32, a pulsar that has been detected in a timing 
anlaysis of COMPTEL data \citep{kuiper1998}.} \end{figure}

\begin{figure}[t!]
\plotone{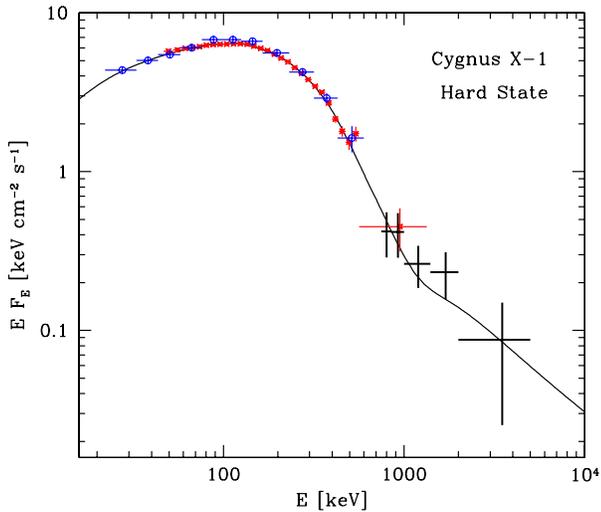}
\caption{The average \gro\/ spectrum of Cygnus X-1 in the hard state fitted
with the {\tt eqpair} model (solid curve). Data points from BATSE and OSSE are 
represented as blue open circles and red asterisks, respectively.  
COMPTEL data are shown as thick crosses.  All
the data are normalized to that of OSSE.  Upper limits have been removed
for the sake of clarity.
\label{f:hard} }
\end{figure}

\begin{figure}[t!]
\plotone{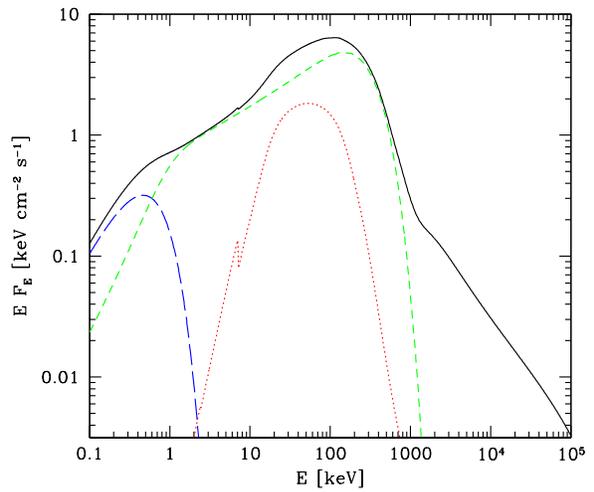}
\caption{Components of the {\tt eqpair} fit for the hard state. All spectra are
intrinsic, i.e., corrected for absorption. The long dashes, short dashes,
and dots correspond to the unscattered blackbody, scattering by
thermal electrons, and Compton reflection, respectively. The solid curve is the
total spectrum. Scattering by the nonthermal electrons accounts for the
high-energy tail above the thermal-Compton spectrum given by the short
dashes, starting at $\sim 1$ MeV.
\label{f:hard_m} }
\end{figure}

\begin{figure}[t!]
\plotone{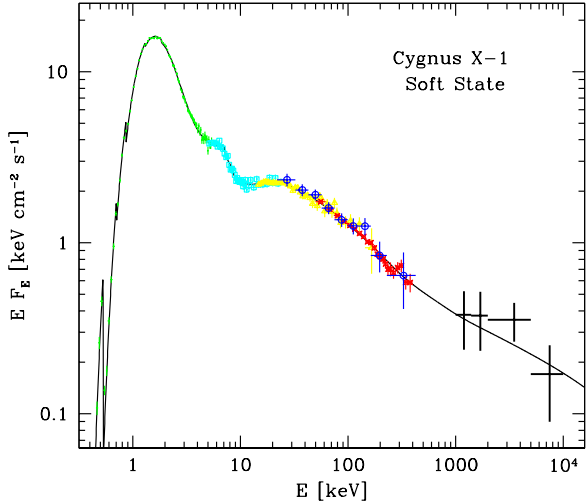}
\caption{The simultaneous \sax-\gro\/ spectrum of
Cygnus X-1 in the soft state fitted with the {\tt eqpair} model (solid curve).
Included are data from the LECS (green), HPGSPC (cyan open squares)
and PDS (yellow open triangles) instruments on board \sax\/ and from
the OSSE (red asterisks), BATSE (blue open circles) and
COMPTEL instruments on \gro\/. All the data are normalized to that of
OSSE.
\label{f:soft} }
\end{figure}

\begin{figure}[t!]
\plotone{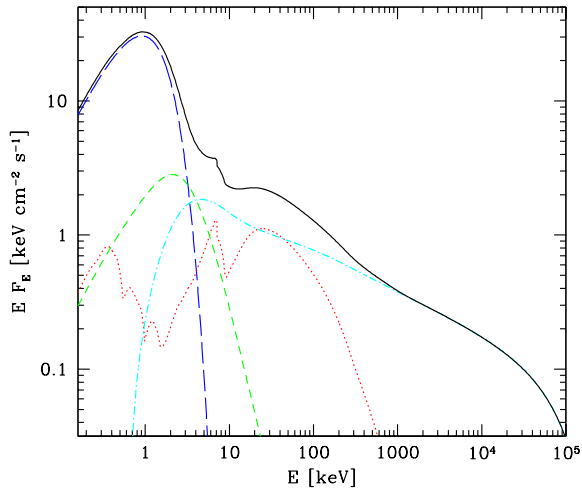}
\caption{Components of the {\tt eqpair} fit for the soft state. All spectra are
intrinsic, i.e., corrected for absorption. The long dashes, short dashes,
dot/dashes and dots correspond to the unscattered blackbody, scattering by
thermal electrons, the scattering by nonthermal electrons, and Compton
reflection/Fe K$\alpha$ fluorescence, respectively. The solid curve is the
total spectrum.
\label{f:soft_m} }
\end{figure}

\begin{figure}[t!]
\plotone{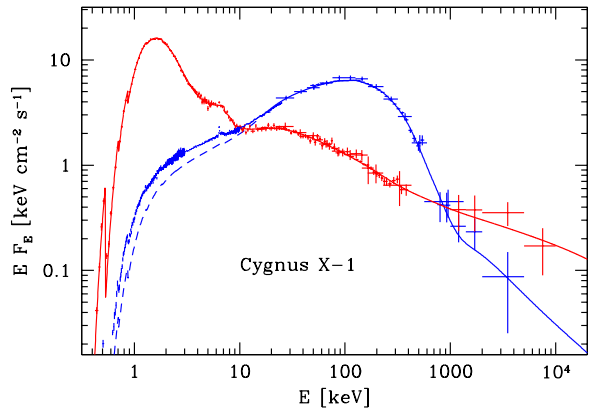}
\caption{Comparison of the spectra in the hard and soft state of Cygnus X-1, 
as fitted with the
{\tt eqpair} model (solid curves). All data are normalized to that of OSSE. The
hard-state data below 25 keV represent a typical hard-state spectrum \citep[the
\sax\/ data of][]{disalvo2001}, and the dashed curve shows the corresponding model
obtained by fitting the \gro\/ data only.
\label{f:compare} }
\end{figure}

\clearpage

\begin{deluxetable}{ccccc} 
\tablewidth{0pt} 
\tablehead{ \colhead{Energy}         
   &  \colhead{Counts in}   & \colhead{Source}      & \colhead{Flux}             
     \\ \colhead{(MeV)}  &  \colhead{Dataspace}   & \colhead{Counts}      &
\colhead{(cm$^{-2}$ s$^{-1}$ MeV$^{-1}$)} } 
\tablecaption{COMPTEL Flux Measurements for VP 522.5.} 
\startdata 
0.75 -- 1.0    &    33,066 & $ 177\pm202$ & $5.4(\pm6.1) \times 10^{-4}$  \\ 
1.0 -- 1.4     &    88,086 & $ 697\pm258$ & $5.3(\pm2.0) \times 10^{-4}$   \\ 
1.4 -- 2.0     &   107,010 & $ 756\pm265$ & $2.6(\pm1.0) \times 10^{-4}$  \\ 
2.0 -- 5.0     &   157,457 & $1092\pm273$ & $5.8(\pm1.5) \times 10^{-5}$  \\ 
5.0 -- 10.0    &    30,158 & $ 191\pm90$  & $5.8(\pm2.7) \times 10^{-6}$  \\ 
10.0 -- 30.0   &     7,967 & $  31\pm31$  & $3.2(\pm3.2) \times 10^{-7}$  \\ 
\enddata 
\end{deluxetable}

\newpage

\begin{deluxetable}{ccccccccccccccc}
\tabletypesize{\scriptsize}
\tablewidth{0pc}
\tablecolumns{15}
\tablehead{ \colhead{$\nh$} & \colhead{$kT_{\rm s}$} & \colhead{$\ls$}
           & \colhead{$\lh/\ls$}  & \colhead{$\lnth/\lh$} & \colhead{$\gamma_{\rm min}$}
           & \colhead{$p$,\,$\Gamma_{\rm inj}$} & \colhead{$kT$} & \colhead{$\tau_{\rm i}$}
           & \colhead{$\tau$} & \colhead{$\Omega/ 2\pi$} 
           & \colhead{$\xi$} & \colhead{$F$} & \colhead{$L$} & \colhead{$\chi^2/\nu$} }
\rotate
\tablecaption{Parameters of the hybrid models for the hard and soft state}
\startdata

\hline
\multicolumn{15}{c}{hard state, {\tt compps}}\\
\hline
&\\
6f & 0.2f & --& --&--& $1.39^{+0.51}_{-0.34}$ & $5.4^{+0.4}_{-0.3}$ &
$58^{+18}_{-22}$ & --& $2.9^{+0.6}_{-0.4}$ & 0.5f &-- & 3.38 & 1.62 & 37/48\\
&\\

\hline
\multicolumn{15}{c}{hard state, {\tt eqpair}}\\
\hline
&\\
6f & 0.2f & $1.8^{+2.5}_{-1.6}$ & $17^{+4}_{-3}$ &
$0.082^{+0.088}_{-0.032}$ &1.5f & $2.0^{+0.9}_{-2.0}$ & 90\tablenotemark{a} &
$1.34^{+0.40}_{-0.50}$ & 1.45\tablenotemark{a} &
$0.52^{+0.06}_{-0.05}$ &-- & 3.56 & 1.70 & 31/46\\

&\\
\hline
\multicolumn{15}{c}{soft state, {\tt compps}}\\
\hline
&\\
$6.0^{+0.2}_{-0.1}$ & $0.39^{+0.01}_{-0.01}$ &-- &-- &-- &
$1.83^{+0.12}_{-0.12}$ &
$3.5^{+0.1}_{-0.1}$ & $63^{+8}_{-8}$ & -- & $0.18^{+0.04}_{-0.02}$ &
$1.4^{+0.3}_{-0.5}$ & $290^{+350}_{-180}$ & 13.1 & 6.3 & 199/239\\
&\\

\hline
\multicolumn{15}{c}{soft state, {\tt eqpair}}\\
\hline
&\\
$6.0^{+0.1}_{-0.1}$  & $0.37^{+0.01}_{-0.01}$ & $3.2^{+38}_{-2.1}$ &
$0.17^{+0.01}_{-0.01}$ &
$0.68^{+0.20}_{-0.12}$ & 1.5f & $2.6^{+0.2}_{-0.2}$ & 65\tablenotemark{a} &
$0.11^{+0.02}_{-0.02}$ & 0.11\tablenotemark{a} &
$1.3^{+0.3}_{-0.3}$ & $100^{+210}_{-60}$ & 13.2 & 6.3 & 199/238\\

&\\

\enddata

\tablecomments{$\nh$, $kT$, and $\xi$ are in units of $10^{21}$ cm$^{-2}$, keV,
and erg\,cm\,s$^{-1}$, respectively. $F$, the unabsorbed model bolometric flux
using the normalization of the OSSE spectrum, and $L$, the corresponding
luminosity assuming isotropy and a distance of 2 kpc are in units of
$10^{-8}$\,erg\,cm$^{-2}$\,s$^{-1}$ and $10^{37}$\,erg\,s$^{-1}$, respectively.
Parameters fixed in the fit are denoted by `f'. The single-parameter
uncertainties correspond to a 90\% confidence, i.e., $\Delta \chi^2=2.71$. }

\tablenotetext{a}{The electron temperature and total optical depth calculated
from energy and pair balance for the best-fit model (i.e., not free
parameters). }

\label{t:pars}
\end{deluxetable}

\end{document}